\documentclass[runningheads]{llncs}
\usepackage[T1]{fontenc}
\pdfoutput=1
\usepackage[linesnumbered,ruled,vlined]{algorithm2e}
\usepackage{listings}
\usepackage{hyperref}
\usepackage{bbding}
\usepackage{float}
\usepackage{xcolor}
\usepackage{amsmath}
\usepackage{multirow}
\usepackage{graphicx}
\newtheorem{defination}{Defination}

\begin{document}
	\title{gMeta: Template-based Regular Expression Generation over Noisy Examples}
	%
	%
	\author{Shujun Wang\inst{1}\orcidID{0000-0001-8161-1695} \and
		Yongqiang Tian\inst{1} \and
		Dengcheng He\inst{1}}
	\authorrunning{Shujun Wang et al.}
	%
	\institute{
		Alibaba Group, Beijing, China\\
		\email{\{wangshujun.wsj,dengcheng.hedc\}@alibaba-inc.com \\puling.tyq@taobao.com}}
	\maketitle              
	\begin{abstract}
		Regular expressions (regexes) are widely used in different fields of computer science, such as programming languages, string processing, and databases. However, existing tools for synthesizing or repairing regexes always assume that the input examples are faultless. In real industrial scenarios, this assumption does not entirely hold. Thus, this paper presents a simple but effective templated-based approach to generate regular expressions over noisy examples. Specifically, we present a data model (i.e., MetaParam) to extract features of strings for clustering all examples. Then, we propose a practical dynamic thresholding scheme to filter out anomalous examples via detecting knee points on CDF graphs. Finally, we design a template-based algorithm to translate a finite of positve examples to regular expression, which is efficient, interpretable, and extensible. We performed an experimental evaluation on four different extraction tasks applied to real-world datasets and obtained promising results in terms of F-measure. Moreover, gMeta achieves excellent results in real industrial scenarios.
		
	\end{abstract}
	

	\section{Introduction}
	A regular expression is a means for specifying string patterns concisely. A specialized engine may use such a specification for extracting the strings matching the specification from a data stream. Regular expressions are a long-established technique for a large variety of text processing applications and continue to be a routinely used tool due to their expressiveness and flexibility\cite{DBLP:conf/emnlp/LiKRVJ08}. Indeed, regular expressions have become an essential device in broadly different application domains, e.g., extraction of bibliographic citations\cite{entity2018} and signal processing hardware design\cite{sidhu2001fast}.
	
	Constructing a regular expression suitable for a specific task is a tedious and error-prone process, which requires specialized skills, including familiarity with the formalism used by practical engines\cite{DBLP:journals/eswa/BarreroRC12}. Multiple approaches have been proposed to generate regular expressions to tackle this problem automatically. The proposal of these works significantly reduces the complexity of learning regular expressions for users. While multiple automatic learning techniques have been proposed, very little work can be applied to the industry directly since they cannot work well with examples containing exceptions. 
	
	Thus, this paper proposes a novel templated-based approach to generate regular expressions over noisy examples. Essential components of our implementation include the following. First, we design an abstract data form MetaParam that can extract example features, which is responsible for extracting the components of a string and the relative order of elements. For example, SMS\_123456 would be translated to MetaParam \textit{X\_d}. Then, we propose an anomaly detection schema based on intelligent threshold setting to filter out MetaParam representing outliers. Crucially, we compute the frequency of each MetaParam to construct a cumulative probability distribution graph. Based on this graph, we can lightly pick the knee point as the anomaly threshold. Finally, we design a template-based regular expression generation scheme, which is efficient, interpretable, and extensible. 
	
	The learning algorithm presented in this work has two significant advantages. First, gMeta is more robust by filtering out abnormal example values. Second, gMeta is no longer committed to directly generating a complex regular expression but instead generates multiple simple regular expressions (depending on the number of MetaParams), significantly reducing the regular expression generation complexity. 
	In summary, the contributions of this work can be outlined as follows:
	\begin{enumerate}
		\item We propose a character-sensitive regular expression template generation scheme, which can capture the common features among multiple strings.
		\item  We design an automatic template-based regular expression generation method gMeta, which can translate a finite of examples containing outliers to regular expressions.
		\item  We formula example outlier filtering problem as a string clustering problem. Crucially, we present a simple and practical approach to extracting string features, namely MetaParam.
		\item We propose an intelligent anomaly threshold generation scheme by computing cumulative probability distribution graphs and knee points.
		\item Extensive experimental results over multiple datasets show the effectiveness of gMeta and a comparison study with the ReLIE algorithm.
	\end{enumerate}
	
	\section{Related Work}
	
	\subsubsection{Programming-by-Example (PBE)} PBE techniques have been the subject of research in the past few decades and successful paradigms for program synthesis, allowing end-users to construct and run new programs by providing examples of the intended program behavior\cite{ref18}. Recently, PBE techniques have been successfully used for string transformations\cite{ref27} , data filtering\cite{ref57} , data structure manipulations\cite{ref60}, table transformations\cite{ref20} , SQL queries\cite{ref56} , MapReduce programs\cite{ref1} , and also regex synthesis\cite{ref3} .
	
	\subsubsection{Regexes Synthesis.} The problem of automatic regex synthesis from examples has been explored in many domains\cite{ref3,ref4}. AlphaRegex\cite{ref32} is an enumeration algorithm for synthesizing simple regexes over binary alphabets from examples. However, all the synthesized expressions are over alphabets of size 2. RegexGenerator $++$\cite{ref3} is a state-of-the-art approach for synthesizing regexes from examples. RegexGenerator++ utilizes genetic programming means that it is not guaranteed to generate a correct solution-i.e., accepting all the positive examples while rejecting all the negative ones. Many existing works focus on XML schemas inference\cite{ref7,ref8}, via resorting to infer regexes from examples. These approaches usually aim to tackle restricted forms of deterministic regexes\cite{ref9} from positive examples only. GP-RegexGolf\cite{ref4}  is an approach based on genetic programming for playing regex golf\cite{ref17} automatically, i.e., for writing the shortest regex that matches all positive strings and does not match any negative string. Unlike many of these efforts, which aim to generalize beyond the provided examples, GPRegexGolf focuses on binary classifying input strings. It does not require a form of generalization, i.e., the ability to induce a general pattern from the provided examples. Several works from the Natural Language Processing community address the problem of generating regexes from natural language specifications based on sequence-to-sequence (seq2seq) model\cite{ref31}.
	
	\subsubsection{Regexes Repair.} several works target repairing or modifying regexes from examples, specifically the incorrect ones. We discuss two main paradigms of them. In the first paradigm, works only consider either positive or negative examples. Li et al.\cite{ref33} proposed ReLIE, which can modify complex regexes by rejecting the newly-input negative examples. By contrast, Rebele et al.\cite{ref44} proposed a novel way to generalize a given regex to accept the given positive examples.
	On the other hand, works in the second paradigm consider both positive and negative examples. Pan et al.\cite{ref39} designed RFixer, a tool for repairing incorrect regexes using both examples. It used skeletons of regexes to prune out the search space effectively, and it employed SMT solvers to efficiently explore the sets of possible character classes and numerical quantifiers. Our work is similar to their work, yet differs in the effectiveness and quality of repaired regexes - we consider not only the correctness but also the ReDoS-invulnerability of the regexes.

	\section{System Design and Implementation} 
	
	\subsection{MetaParam}
	This subsection proposes a MapReduce-based Parameter abstraction approach, extracting the common features of an ocean of examples. Based on this, we can cluster examples and then compute the frequency of each MetaParam.
	
	\IncMargin{1em}
	\begin{algorithm} \SetKwData{Left}{left}\SetKwData{This}{this}\SetKwData{Up}{up} \SetKwFunction{Union}{Union}
		\SetKwFunction{ExtractCommonSubsequence}{ExtractCommonSubsequence}
		\SetKwFunction{TransformAndCompress}{TransformAndCompress}
		\SetKwFunction{LengthComputing}{LengthComputing}
		\SetKwFunction{MergeSubsequence}{MergeSubsequence}
		\SetKwFunction{MergePatterns}{MergePatterns}
		\SetKwFunction{MergeLengths}{MergeLengths}
		\SetKwInOut{Input}{Input}\SetKwInOut{Output}{Output}
		
		\Input{A set of examples $V = v_1 \cup v_2 \cup \ldots \cup v_n$} 
		\Output{A common subsequence $s$ \\ A set of example patterns $P = p_1 \cup \ldots \cup p_n  $ }
		\BlankLine 
		
		\textbf{Mapper}
		
		\qquad $s_i \leftarrow$\ExtractCommonSubsequence{$v_i$}\; 
		\qquad $p_i \leftarrow$\TransformAndCompress{$v_i$}\; 
		
		\textbf{Reducer}
		
		\qquad $s \leftarrow$\MergeSubsequence{$s_i$}\;
		\qquad $P \leftarrow$\MergePatterns{$p_i$}\;
		\caption{Example Abstraction}
		\label{metaparam} 
		\Return $s$, $P$
	\end{algorithm}
	
	As shown in Algorithm~\ref{metaparam}, we take a two-stage approach: Mapper and Reducer, to extract common example features from an ocean of values. We emphasize two types of example features: character analysis and character relative sequence analysis.	There are two main stages of the example abstraction:
	
	\textbf{Mapper:} In the mapper stage, we mainly focus on extracting the local features of each example. Specifically,
	\begin{itemize}
		\item \textbf{Extracting Common Subsequence} will extract the longest common subsequences in the example set $V$.
		\item \textbf{Transformer And Compress} will translate a specific example into an abstract form. The translation rules are as follows(See Table~\ref{tab:transform}):
		\begin{table}
			\centering
			\caption{Transform Rules\label{tab:transform}}
			\begin{tabular}{l|l} 
				\hline
				Representation & Description \\
				\hline  
				\textit{z} & The Chinese character \\
				\textit{x} & English lowercase characters  \\
				\textit{X} & English uppercase characters \\
				\textit{d} & Number  \\
				& Other characters reserved\\
				\hline
			\end{tabular}
		\end{table}
		The transformer stage is responsible for abstract examples. For instance, the number $123$ will be abstracted as $ddd$, and the compressing stage will further translate it into a character $d$.
	\end{itemize}

	\textbf{Reducer:} In this stage, the local example features are combined to extract the global features of examples. Crucially,  we merge the MetaParams produced in the Mapper stage and compute the frequency of each MetaParam.
	\subsection{Outlier Filtering}
	To efficiently filter out outliers in massive examples, gMeta adopts a bottom-up strategy at the MetaParam level to shrink the solution space. More specifically, gMeta filters out most low-frequency MetaParam and then translate the remaining potentially normal examples into regular expressions.
	\begin{figure}[!h]
		\centering
		\includegraphics[width=0.5\linewidth]{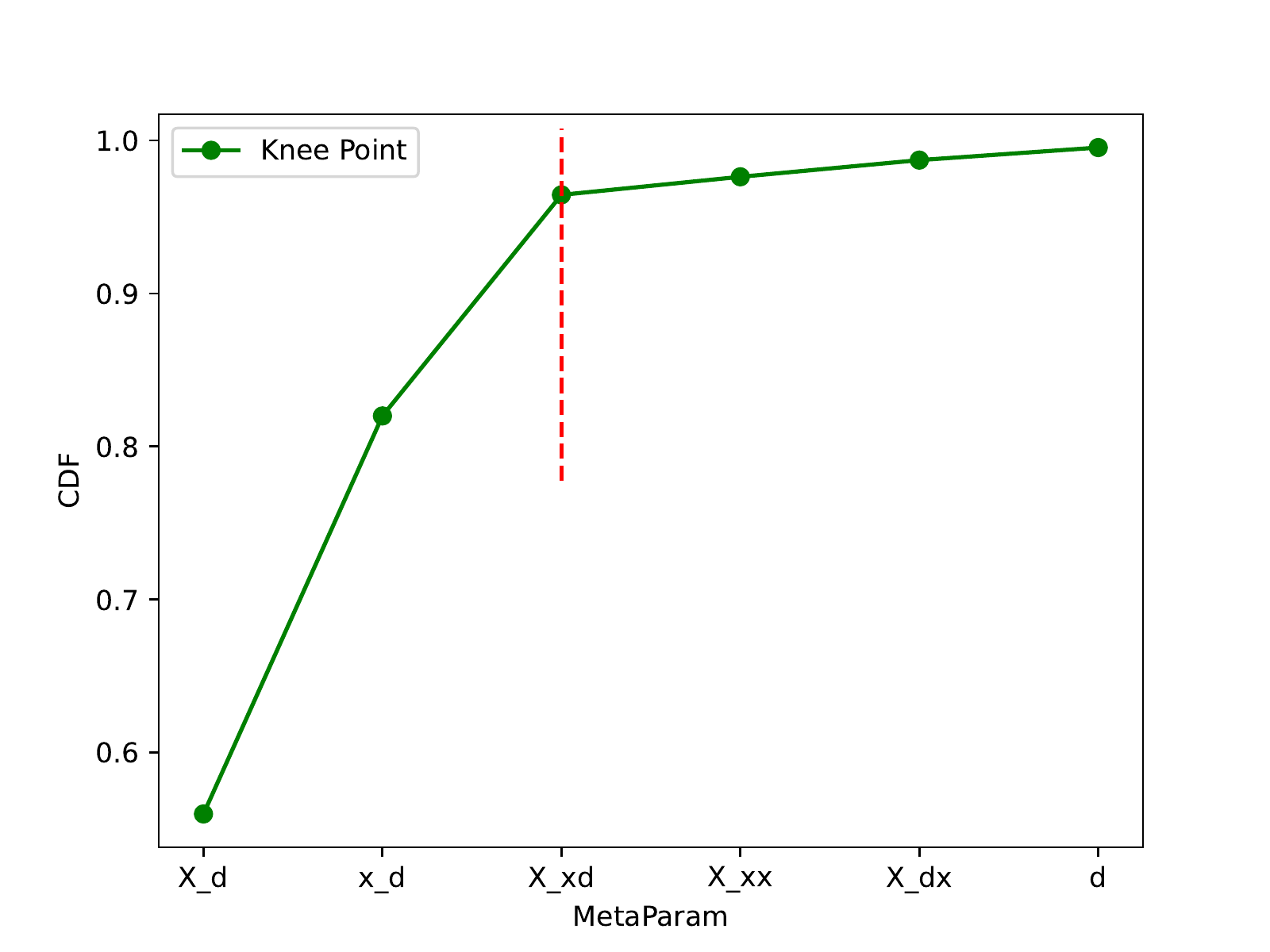}
		\caption{Cumulative distribution of the frequency of MetaParam, and knee point threshold selection.}
		\label{knee}
	\end{figure}
	
	\begin{theorem}\label{tm1}
		Given a large number of examples, the number of "abnormal" examples in real industrial scenarios is usually much less than those of "normal" ones. 
	\end{theorem}
	Similar to those anomaly detection algorithms, we apply a threshold to decide whether the MetaParam is an anomaly or not. Figure~\ref{knee} shows the cumulative distribution of MetaParam frequencies. The skewed distribution naturally allows us to apply the knee-point method to pick the knee-point as the anomaly threshold automatically. For example, the vertical dashed line in Figure~\ref{knee} marks the threshold the knee method gives for this example.
	
	As shown in Figure~\ref{knee}, examples of abstract form X\_d, x\_d and x\_xd will be parsed as regular expressions, respectively. Note that here we generate a regular expression for each MetaParam.
	\subsection{Regular Expression Generation}
	Given a set of normal examples, this stage will generate a finite of highly available regular expressions.
	\subsubsection{Template Generation} As shown in Figure~\ref{template}, the template mainly comprises two types of elements, \textit{slot} and \textit{character} (i.e., the longest common subsequence). We extract the longest common subsequence in multiple normal examples to serve as template anchors, which can effectively capture common character features and relative position information between strings. Then, a variable component \textit{slot} will be added to the middle of all non-consecutive common characters.
	
	\subsubsection{Slot Generation}
	\begin{figure}[htbp]
		\centering
		\begin{minipage}[t]{0.55\textwidth}
			\centering
			\includegraphics[width=\linewidth]{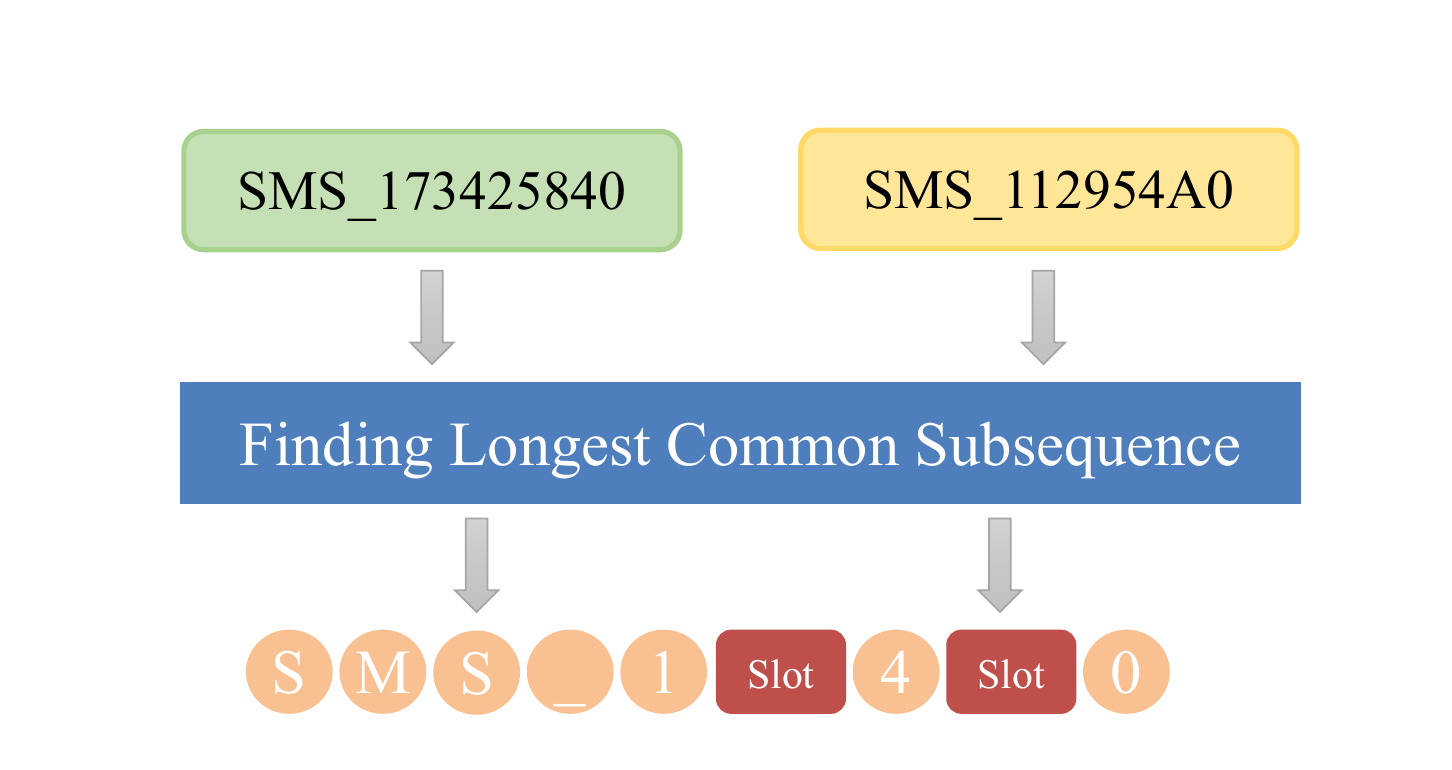}
			\caption{An example of template generation.}
			\label{template}
			
		\end{minipage}
		\begin{minipage}[t]{0.44\textwidth}
			\centering
			\includegraphics[width=\linewidth]{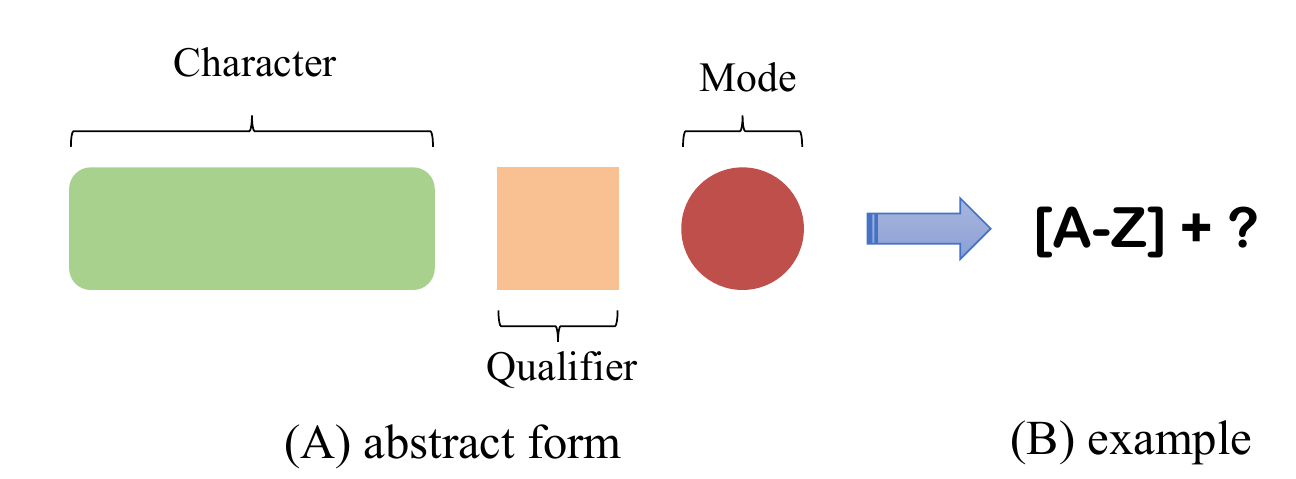}
			\caption{Slot form.}
			\label{slot}
		\end{minipage}
	\end{figure}
	
\begin{figure}[!ht]
		\centering
		\includegraphics[width=0.88\linewidth]{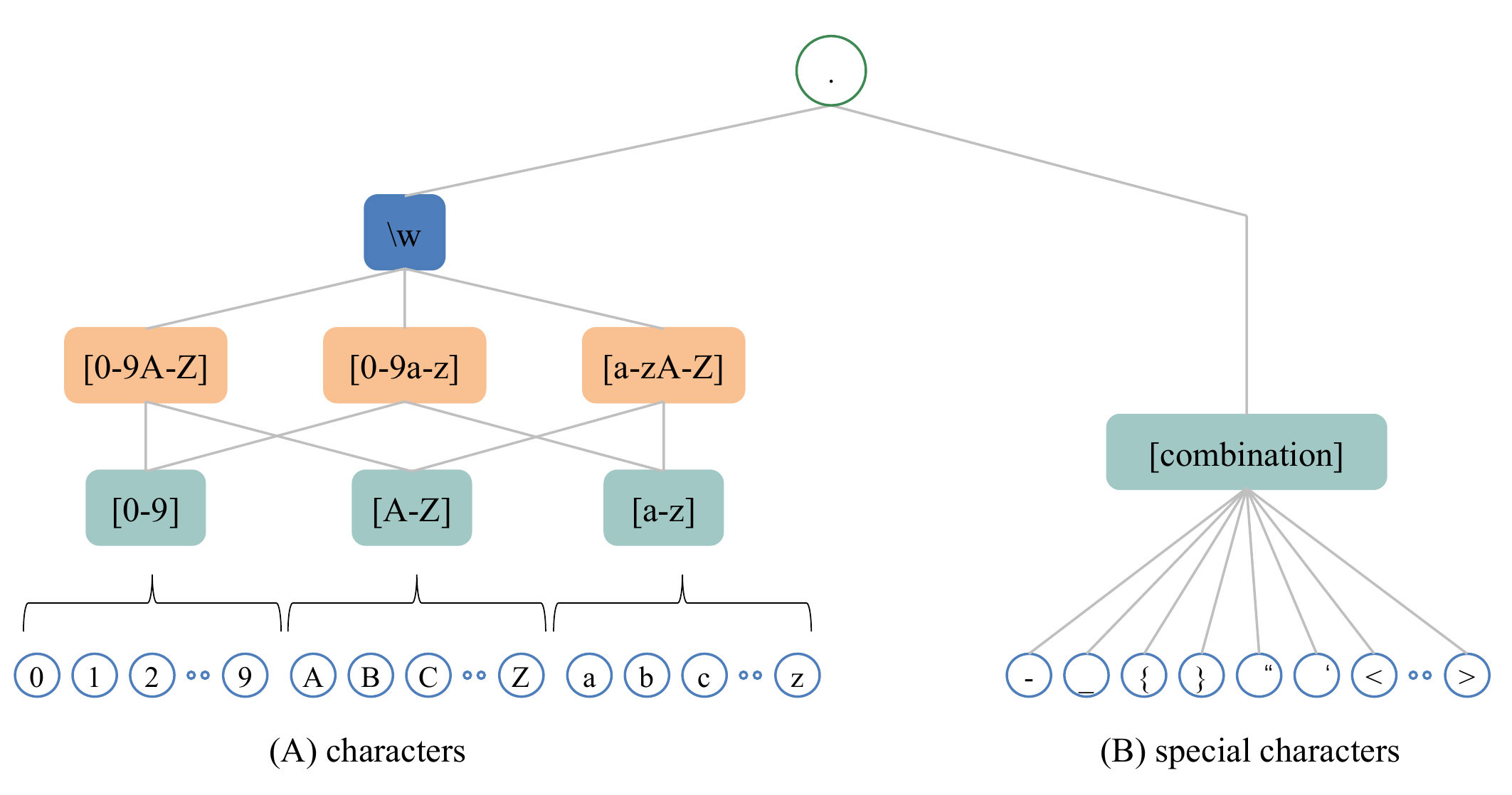}
		\caption{Hierarchical Abstract Tree.}
		\label{regular}
	\end{figure}

	Figure~\ref{slot}(A) exhibits the abstract slot form, and Figure~\ref{slot}(B) is a slot example. Characters are short-hand notations for denoting the disjunction of a set of characters ( $\backslash d$ is equivalent to $(0|1 \ldots| 9)$.
	Quantifiers are used to define the range of valid counts of a repetitive sequence. For instance, $a\{m, n\}$ looks for a sequence of a's of length at least $m$ and at most $n$. Mode indicates the matching mode; for example, character $?$ denotes a non-greedy match.
	The regex learning task addressed in this subsection can now be formally stated as the following optimization problem:
	\begin{defination}
		Given a finite of strings, how to generate an appropriate regular expression, i.e., slot in this paper.
	\end{defination}
	\begin{figure}[!ht]
		\centering
		\includegraphics[width=\linewidth]{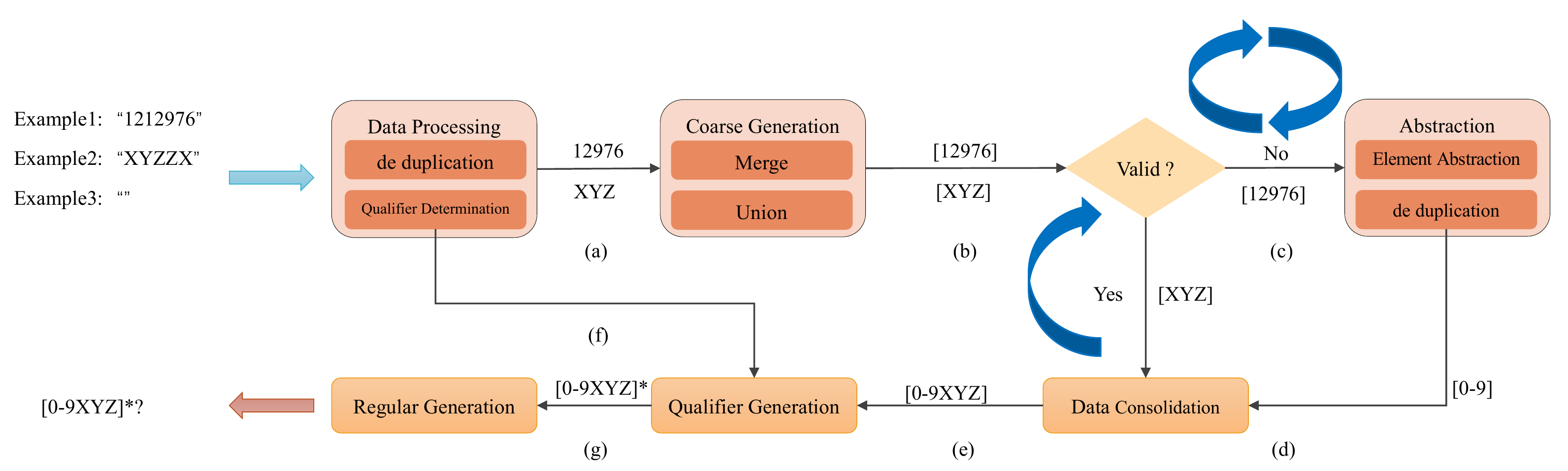}
		\caption{A Running Example of Slot Generation.}
		\label{gMetaExample}
	\end{figure}	
	Figure~\ref{gMetaExample} exhibits a running example of slot generation, and the steps are as follows:
	\begin{itemize}
		\item The data processing module mainly plays two roles; one is to deduplicate the example. The second is to compute the length information of the string and pass the length information to the module \textit{Qualifier Generation}.
		\item The coarse-grained regular expression generation module generates a regular expression via adding special characters "[" and "]" on both sides of the string produced by the data processing module.
		\item The valid module determines whether the generated coarse-grained regular expression is valid. In this paper, we limit the number of characters in the regular expression(i.e., The number of characters is less than 4).
		\item The abstraction module further abstracts coarse-grained regular expressions $p_i$ that do not satisfy predefined rules. As shown in Figure~\ref{regular}, the characters at the level closest to the leaf node in $p_i$ are abstracted upward (i.e., the parent node level). For instance, character 0 will be abstracted as its parent nodes [0-9], and the further abstracted regular expressions will be deduplicated. If the results still do not satisfy predefined rules, the abstract process will be repeated until they meet the requirements. 
		\item The data consolidation module is responsible for merging multiple single regular expressions.
		\item The qualifier generation module generates qualifiers according to the length information of the strings. For example, if there is an empty string, the qualifier will be set to "*".
		\item The regular expression generation module will wrap the final result, such as adding "/\ " for special characters "." .
	\end{itemize}

	\section{Evaluation}
	This section presents an empirical study of the gMeta algorithm using four extraction tasks over three real-life data sets and a real Industrial scene dataset. This study aims to evaluate the effectiveness of gMeta and investigate how it compares with existing automatic regular expression generation work.
	
	\subsection{Experimental Setup}
	\subsubsection{Datasets} We use 4 datasets presented in ReLIE\cite{ReLIE08} to validate the effectiveness of gMeta. The ReLIE dataset\footnote{http://dbgroup.eecs.umich.edu/project/regexLearning/} includes four tasks (phone numbers, course numbers, software names, and URLs). Each task comes with a set of documents. Each document contains a span of words and 100 characters of context to the left and right. Each span is annotated as a positive or a negative example. We manually cleaned the dataset by fixing obvious annotation errors, e.g., where a word is marked as a positive and a negative example in the same task. In total, the dataset contains 90807 documents. Furthermore, we construct a dataset named AliRegex, which contains an ocean of examples containing outliers extracted from API call logs.
	
	\subsubsection{Measures}
	We evaluated the precision, recall, and F-measure of gMeta on multiple datasets. In detail, we count an extraction when some (non-empty) string has been extracted from an example and a correct extraction when exactly the (non-empty) string associated with a positive example has been extracted. Accordingly, the precision of a regular expression is the ratio between the number of correct extractions and the number of extractions. The recall is the ratio between the number of correct extractions and the number of positive examples; F-measure is the harmonic mean of precision and recall. Note that the formula for computing precision on datasets containing anomalous examples is as follows:
	\begin{equation}
		Prec = \dfrac{||Correct~Extractions||- ||Outliers~Extraction||}{||Normal~Examples||}
	\end{equation}
	\subsection{Results}
	\begin{table}[!h]
		\centering
		\begin{tabular}{|l|ccc|c|}
			\hline \multirow{2}{*}{ Task } &
			\multicolumn{3}{|c|}{ Dataset } & { Results (\%) }  \\
			& Learning & $\%$ & Testing & F-measure   \\
			\hline & 25 & $5$ & 12 & $77.8$   \\
			Phone Number & 50 & $10$ & 25 & $86.4$   \\
			& 100 & $20$ & 50 & $88.9$   \\
			\hline & 25 & $5$ & 12 & $9.2$  \\
			Course Number & 50 & $10$ & 25 & $9.6$   \\
			& 100 & $20$ & 50 & $10.1$   \\
			\hline 
			\multirow{3}{*}{ Software Name } & 25 & $5$ & 12 & $54.5$   \\
			& 50 & $10$ & 25 & $67.2$  \\
			& 100 & $20$ & 50 & $77.6$   \\
			\hline
			\multirow{3}{*}{ Urls } & 25 & $5$ & 12 & $74.5$   \\
			& 50 & $10$ & 25 & $82.3$  \\
			& 100 & $20$ & 50 & $87.6$   \\
			\hline
		\end{tabular}
		\caption{Experiment results with different datasets \label{tab:result}}
	\end{table}
	
	Let \textit{Learning} in Table~\ref{tab:result} denotes how many examples are used to generate the regular expression. For example, \textit{Learning} 12 means that gMeta will randomly sample 12 examples to generate a regular expression. We executed an extensive suite of experiments by varying the size of the learning set, as summarized in Table~\ref{tab:result}. The performance of gMeta on the four datasets varies widely. From the perspective of F1 indicators, the order from high to low is: 
	\begin{equation}
		Urls \approx Phone~Number > Software~Name \gg Course~Number.
	\end{equation}
	
	We find that gMeta has significant scene applicability. Specifically, gMeta usually works well if the correct expression contains fixed characters, special characters, or specific length information. However, if the example does not contain some implicit pattern, gMeta's performance is not as good as it should be.
	\begin{table}[!ht]
		\centering
		\begin{tabular}{|c|c|c|c|c|c|c|}
			\hline \multirow{2}{*}{ } & \multicolumn{2}{|c|}{$5 \%$} & \multicolumn{2}{c|}{$10 \%$} & \multicolumn{2}{c|}{$20 \%$} \\
			\cline { 2 - 7 } & ~gMeta~ &~ReLIE~& ~gMeta~ & ~ReLIE~ & ~gMeta~ & ~ReLIE~ \\
			\hline \hline AliRegex & $\mathbf{0 . 645}$ & $0.396$ & $\mathbf{0 . 722}$ & $0.501$ & $\mathbf{0 . 816}$ & $0.624$ \\
			\hline
		\end{tabular}
		\caption{gMeta VS ReLIE\label{tab:vs}}
	\end{table}
	
	As shown in Table\ref{tab:vs}, we compare the effectiveness of gMeta and ReLIE on datasets containing outliers, gMeta performs significantly better than ReLIE.
	
	\section{Conclusions}
	We have proposed a template-based approach for the automatic generation of regular expressions. The approach requires only a set of examples to describe the extraction task and does not require any hint about the regular expression that solves that task. Users thus require no specific skills in regular expressions.
	We assessed our proposal on three datasets from different application domains. The results in terms of precision are promising, even if compared to earlier state-of-the-art proposals ReLIE. The execution time is sufficiently short of making the approach practical.
	Crucially, different from previous research on automatic regular expression generation, gMeta no longer assumes that all input examples are correct but considers the existence of outliers, making gMeta more robust and scalable. The template-based generation method allows users to make targeted adjustments to make the generated regular expressions more aligned with industrial requirements. In addition, gMeta does not seek to cover all cases with one regular expression. Usually, gMeta will generate several regular expressions. In this case, we can significantly reduce the complexity of a single regular expression. Although the number of regular expressions has increased, their execution efficiency is still very optimistic in the current case of parallel computing.
	
	%
	%
	\section{Acknowledgments} 
	This work was supported by Alibaba Group through Alibaba Innovative Research Program.
	\bibliographystyle{splncs04}
	\bibliography{reference}

\end{document}